\documentclass[final,5p,times,twocolumn]{elsarticle}

\usepackage{bm}
\usepackage{bbm}
\usepackage{bbold}
\usepackage{slashed}
\usepackage{verbatim}
\usepackage{cancel}
\usepackage{graphicx}
\usepackage{combelow} 
\usepackage{mathtools}
\usepackage[usenames,dvipsnames,svgnames,table]{xcolor}
\usepackage[colorlinks=True,linkcolor=blue,citecolor=blue,urlcolor=blue]{hyperref}
\usepackage{soul}
\usepackage{ulem}

\let\oldciteauthor=\citeauthor
\def\citeauthor#1{\hypersetup{citecolor=black}\oldciteauthor{#1}}

\let\oldciten=\onlinecite
\def\onlinecite#1{\hypersetup{citecolor=blue}\oldciten{#1}}

\let\oldcite=\cite
\def\cite#1{\hypersetup{citecolor=blue}\oldcite{#1}}

\makeatletter
\newsavebox{\@brx}
\newcommand{\llangle}[1][]{\savebox{\@brx}{\(\m@th{#1\langle}\)}%
  \mathopen{\copy\@brx\kern-0.5\wd\@brx\usebox{\@brx}}}
\newcommand{\rrangle}[1][]{\savebox{\@brx}{\(\m@th{#1\rangle}\)}%
  \mathclose{\copy\@brx\kern-0.5\wd\@brx\usebox{\@brx}}}
\makeatother

\allowdisplaybreaks

\journal{Physics Letters B}

\begin{document}

\begin{frontmatter}

\title{Thermodynamics of rotating fermions}

\author[uvt]{Victor E. Ambru\cb{s}}
\ead{victor.ambrus@e-uvt.ro}
\author[uvt]{Aleksandar Geci\'{c}}

\address[uvt]{Department of Physics, West University of Timi\cb{s}oara,\\
Bd.~Vasile P\^arvan 4, Timi\cb{s}oara 300223, Romania}

\begin{abstract}
We consider the thermodynamic properties of a rotating gas of free fermions. We begin by constructing the thermodynamic potential $\Phi$ and its associated current $\phi^\mu$ within the grand canonical ensemble of a macroscopic rigidly rotating body, where the ensemble parameters are the temperature $T_0$ and chemical potential $\mu_0$ on the rotation axis, as well as the rotation angular velocity $\Omega_0$. We then consider the problem of local thermodynamics, where the thermodynamic state is defined by the local temperature $T$ and chemical potential $\mu$, as well as the local spin potential tensor, $\Omega_{\mu\nu}$. We find the thermodynamic pressure $P$, given as the sum of the usual classical (non-quantum) pressure and other corrections due to the spin potential and the kinematic state of the fluid, based on the local expectation value of the energy-momentum tensor. We compute the associated entropy, charge and spin densities, and show they are consistent with the Euler relation. 
\end{abstract}

\end{frontmatter}

\date{\today}

\section{Introduction}

Quantum systems under rotation have been studied for almost half of a century. In the late '70s, Vilenkin \cite{Vilenkin:1978is,Vilenkin:1980zv, Vilenkin:1978hb, Vilenkin:1979ui} showed that a gas of neutrinos in thermodynamic equilibrium, undergoing rotation, develops a flow in the direction parallel to the angular velocity. This phenomenon, later named the axial vortical effect \cite{Kharzeev:2015znc}, states that the Dirac field under rotation exhibits a flow of axial charge, $\mathbf{j}_A = \sigma^\omega_A \boldsymbol{\omega}$, parallel to the vorticity vector. The proportionality factor, the axial vortical conductivity $\sigma^\omega_A = T^2 / 6 + \mu^2 / 2\pi^2$, can be related to the triangle diagrams characterizing the anomalous non-conservation of the axial current in an interacting theory \cite{Landsteiner:2011cp}.

Rigid rotation can be treated unproblematically for a classical gas \cite{Cercignani:2002}. At the quantum level, expectation values computed over the full set of quantum modes supported in infinite spacetime invariably involve ``superhorizon modes'', whose wavelengths stretch beyond the light cylinder, where a corotating observer reaches the speed of light. For fermions, this leads to a discrepancy between the static and rotating vacua \cite{Iyer:1982ah,Ambrus:2014uqa}. In the case of bosons, the consequence is more dramatic: modes with vanishing corotating energy, $\widetilde{E} = E - \Omega_0 m$, expressed as the difference between the static energy $E$ and the product between the angular momentum $m$ of the mode and the angular velocity $\Omega_0$, exhibit a divergent distribution, $[e^{\beta \widetilde{E}} - 1]^{-1} \to \infty$. This makes rigidly rotating states of bosons ill defined \cite{Vilenkin:1980zv,Frolov:1989jh,Duffy:2002ss}.
The causality issue and the aforementioned consequences can be avoided by imposing boundary conditions. However, boundary-related complications make analytical treatments of the system untractable \cite{Ambrus:2015lfr,Singha:2024tpo,Singha:2025zvh}. 

Despite the above issues, thermal expectation values can be computed for a fermion gas everywhere inside the light cylinder \cite{Ambrus:2014uqa}. In the case of the scalar field, expectation values can be computed either perturbatively, around the static equilibrium \cite{Becattini:2015nva}, or by using various techniques, such as analytic distillation \cite{Becattini:2020qol} or the non-perturbative approach in Ref.~\cite{Ambrus:2017opa}. These techniques reveal quantum corrections that extend the classical expressions derived in, e.g., kinetic theory \cite{Ambrus:2017opa}. The thermodynamic consistency of these results has been a long-standing problem \cite{Becattini:2013fla}. As pointed out in Refs.~\cite{Becattini:2023ouz,Becattini:2025oyi}, it is a non-trivial task to find an expression for the thermodynamic pressure that simultaneously satisfies the Euler relation, relating it to the energy, entropy, charge and spin densities, as well as the differential relations inherited from the thermodynamic potential.

The purpose of this paper is to provide a rigorous and complete analysis of the rigidly rotating system, to serve as the basis for the formulation of the thermodynamics of fluids with spin and vorticity. 
We focus on the expectation values of the energy-momentum tensor $\widehat{T}^{\mu\nu}$ and the vector $\widehat{J}^\mu_V$ and axial $\widehat{J}^\mu_A$ charge currents, which are quadratic operators with respect to the field operator $\hat{\psi}$. We leave the discussion of higher-order operators, relevant for, e.g., interacting theories, for future work.
We begin with Sec.~\ref{sec:TEV}, where we review the thermal expectation values of the energy-momentum tensor and charge currents derived in quantum field theory under rotation (see Refs.~\cite{Ambrus:2019ayb,Ambrus:2019khr} for details). Within the grand canonical ensemble (GCE), we construct the grand potential and the associated thermodynamic potential current \cite{Becattini:2023ouz} in Sec.~\ref{sec:GCE}. These quantities, constructed in the GCE, are exact and fully thermodynamically consistent. The transition to the local description of the vortical fluid is made in Sec.~\ref{sec:local}, where we start with a discussion of exchanging the parameters of the GCE (global temperature $T_0$, chemical potential $\mu_0$ and angular velocity $\Omega_0$) with local ones. We also emphasize the importance to distinguish between the vorticity tensor $\omega_{\alpha\beta}$ and the spin potential, $\Omega_{\alpha\beta}$. We finally provide a resolution to the problem of local thermodynamics for the exactly solvable case of massless fermions. Our conclusions are presented in Sec.~\ref{sec:conc}.

\section{Quantum expectation values}\label{sec:TEV}

We consider a thermodynamic state of free fermions, distributed according to the density operator \cite{Becattini:2012tc,Salvio:2025ggj}
\begin{equation}
 \hat{\rho} = e^{-\beta_0 (\widehat{H} - \mu_0 \widehat{Q} - \boldsymbol{\Omega}_0 \cdot \widehat{\mathbf{J}})},
 \label{eq:rho}
\end{equation}
where $\widehat{H}$ is the Hamiltonian, $\widehat{Q} \equiv \widehat{Q}_V$ is the conserved (vector) charges and $\boldsymbol{\Omega}_0 \cdot \widehat{\mathbf{J}}$ is the total angular momentum projected along the angular velocity vector $\boldsymbol{\Omega}_0 = \Omega_0 \mathbf{e}_z$, taken, without loss of generality, along the vertical axis. 

The thermal expectation value of an operator $\widehat{A}$ is $\langle\widehat{A} \rangle = Z^{-1} {\rm tr}(\hat{\rho} \widehat{A})$, where $Z = {\rm tr}(\hat{\rho})$ is the partition function and the trace is taken over the entire Fock space. The field operator $\hat{\psi}(x)$ is expanded with respect to particle modes as 
\begin{equation}
 \hat{\psi}(x) = \sum_j [\theta(\sigma_j) U_j(x) \hat{a}_j + 
 \theta(-\sigma_j) V_j(x) \hat{b}^\dagger_j],
\end{equation}
where the antiparticle modes $V_j$ are related to the particle modes via charge conjugation, $V_j(x) = i \gamma^2 U_j^*(x)$. The associated modes $U_j$ and $V_j$ are well-known \cite{Ambrus:2014uqa,Ambrus:2019cvr} and are not repeated here, for brevity. The index $j$ collects all eigenvalues defining the cylindrical modes, i.e. $j = \{E_j, k_j, m_j, \lambda_j, \sigma_j\}$. These quantities (energy $E_j$; vertical momentum $k_j$; vertical angular momentum $m_j = \pm \frac{1}{2}$; $\pm \frac{3}{2}$, $\dots$; helicity $\lambda_j = \pm 1/2$; and particle charge $\sigma_j = \pm 1$) correspond to the system of commuting operators defining the individual (anti-)particle modes \cite{Ambrus:2019cvr}, i.e.
\begin{equation}
 [\widehat{H}, \hat{a}^\dagger_j] = E_j \hat{a}^\dagger_j, \quad 
 [\widehat{P}^z, \hat{a}^\dagger_j] = k_j \hat{a}^\dagger_j, \quad 
 [\widehat{J}^z, \hat{a}^\dagger_j] = m_j \hat{a}^\dagger_j,
\end{equation}
and similar for antiparticles, while 
$[\widehat{Q}_V, \hat{a}^\dagger_j] = \hat{a}^\dagger_j$ and $[\widehat{Q}_V, \hat{b}^\dagger_j] = -\hat{b}^\dagger_j$.

In this work, we only consider operators which are quadratic in the field operator $\widehat{\psi}$, whose expectation values can be taken with respect to the product of one-particle operators for cylindrical modes:
\begin{equation}
 \langle \hat{a}^\dagger_j \hat{a}_{j'} \rangle = \left.f_j \delta(j, j')\right|_{\sigma_j = 1}, \quad 
 \langle \hat{b}^\dagger_j \hat{b}_{j'} \rangle = \left.f_j \delta(j, j')\right|_{\sigma_j = -1},
\end{equation}
where $\delta(j,j') = E_j^{-1} \delta(E_j - E_{j'}) \delta(k_j - k_{j'}) \delta_{m_j,m_{j'}} \delta_{\lambda_j, \lambda_{j'}} \delta_{\sigma_j, \sigma_{j'}}$ and 
\begin{equation}
 f_j = \left[e^{\beta_0(E_j - \Omega_0 m_j - \sigma_j \mu)} + 1\right]^{-1}. 
 \label{eq:fj}
\end{equation}

We will consider the expectation values of the (canonical) energy-momentum tensor $\widehat{\Theta}^{\mu\nu}$, vector current $\widehat{J}^\mu_V$ and axial current $\widehat{J}^\mu_A$, defined as
\begin{gather}
 \widehat{\Theta}^{\mu\nu} = \frac{i}{2} \hat{\bar{\psi}} \gamma^\mu \overleftrightarrow{\partial}^\nu \hat{\psi}, \quad
 \widehat{J}^\mu_V = \hat{\bar{\psi}} \gamma^\mu \hat{\psi}, \quad 
 \widehat{J}^\mu_A = \hat{\bar{\psi}} \gamma^\mu \gamma^5 \hat{\psi}.
\end{gather}
Generically, the expectation value  $A \equiv \langle \widehat{A} \rangle$ of an operator $\widehat{A}$ reads
\begin{equation}
 A= \frac{1}{8\pi^2} \sum_j f_j \mathcal{A}_j,
 \label{eq:A}
\end{equation}
where $\sum_j = \sum_{\sigma,\lambda,m} \int_M^\infty dE\, E \int_{-p}^p dk$, with $M$ being the fermion mass. The sesquilinear forms $\mathcal{A}_j$ appearing above were derived in Ref.~\cite{Ambrus:2019ayb} and are repeated here without derivations. For the vector current, we have
\begin{gather}
 \mathcal{J}^t_{V;j}
  = \sigma_j\left(J_j^+ + \frac{2\lambda_j k_j}{p_j} J_j^-\right), \quad 
 \mathcal{J}^\varphi_{V;j} = \frac{\sigma_j q_j}{E_j} J_j^\times, \nonumber\\
 \mathcal{J}^z_{V;j} = \sigma_j \left(\frac{k_j}{E_j} J_j^+ + \frac{2\lambda_j p_j}{E_j} J_j^-\right),
\end{gather}
while for the axial current,
\begin{gather}
 \mathcal{J}^t_{A;j} = \frac{2\lambda_j p_j}{E_j} J_j^+ + \frac{k_j}{E_j} J_j^-, \quad 
 \mathcal{J}^\varphi_{A;j} = \frac{2\lambda_j q_j}{p_j} J_j^\times, \nonumber\\
 \mathcal{J}^z_{A;j} = J_j^- + \frac{2\lambda_j k_j}{p_j} J_j^+,
\end{gather}
where we introduced the notation
\begin{align}
 J_j^\pm &= J_{m_j - \frac{1}{2}}^2(q_j \rho) \pm J_{m_j + \frac{1}{2}}^2(q_j \rho), \nonumber\\
 J_j^\times &= 2 J_{m_j - \frac{1}{2}}(q_j \rho) J_{m_j + \frac{1}{2}}(q_j \rho),
\end{align}
where $J_{m_j \pm \frac{1}{2}}(q_j \rho)$ are Bessel functions of the first kind.
For the energy-momentum tensor, we have 
\begin{gather}
 \Theta_j^{\mu t} = E_j \sigma_j \mathcal{J}^\mu_{V;j}, \quad 
 \Theta_j^{\mu z} = k_j \sigma_j \mathcal{J}^\mu_{V;j}, \nonumber\\
 \Theta^{\rho\rho}_j = \frac{q_j^2}{E_j} J_j^+ - \frac{q_j m_j}{\rho E_j} J_j^\times, \quad 
 \Theta^{\varphi\varphi}_j = \frac{m_j q_j}{\rho^3 E_j} J_j^\times, \nonumber\\
 \rho^2 \Theta_j^{t\varphi} = m_j J_j^+ + \frac{2\lambda_j k_j m_j}{p_j} J_j^- - \frac{1}{2} \mathcal{J}^z_{A;j}, \nonumber\\ 
 \rho^2 \Theta_j^{z\varphi} = \frac{k_j m_j}{E_j} J_j^+ + \frac{2\lambda_j p_j m_j}{E_j} J_j^- - \frac{1}{2} \mathcal{J}^t_{A;j}.
\end{gather}
We omit the off-diagonal components involving one radial index, i.e. $\Theta^{t\rho}$, $\Theta^{\rho\varphi}$, $\Theta^{\varphi\rho}$ and $\Theta^{z\rho}$, as these components play no role in the following discussion.
Under the mode sum in Eq.~\eqref{eq:A}, the terms which are odd with respect to $k \to -k$ or $\lambda \to -\lambda$ will cancel, due to the fact that the distribution function $f_j$ in Eq.~\eqref{eq:fj} is even under these transformations. We will nevertheless continue displaying these terms, without taking into account the symmetries of $f_j$ (other than the fact that $\hat{\rho}$ is diagonal in the chosen cylindrical basis), in order to keep our statements as general as possible.

\section{Grand canonical ensemble}\label{sec:GCE}

\subsection{Grand potential}\label{sec:GCE:GP}

In this section, we discuss the properties of the grand canonical ensemble, defined by the density operator \eqref{eq:rho}. To preserve causality, the system must be enclosed within a cylindrical boundary of radius $R \le \Omega_0^{-1}$, on which suitable boundary conditions must be employed \cite{Ambrus:2015lfr}. Because this leads to complications in the particle spectrum, we instead consider that the system is contained in a fictitious cylinder of radius $R$, without imposing boundary conditions, and construct the grand canonical potential $\Phi$ (usually defined by $\Phi = -T_0 \ln Z$) given as 
\begin{equation}
 \Phi = \int_V d^3x \phi(x).
\end{equation}
By definition, the grand canonical potential satisfies
\begin{equation}
 \Phi = \mathcal{E} - T_0 \mathcal{S} - \mu_0 \mathcal{Q} - \boldsymbol{\Omega}_0 \cdot \boldsymbol{\mathcal{M}},
\end{equation}
where the total entropy, charge and angular momentum are obtained via
\begin{equation}
 \mathcal{S} = -\frac{\partial \Phi}{\partial T_0}, \quad 
 \mathcal{Q} = -\frac{\partial \Phi}{\partial \mu_0}, \quad 
 \boldsymbol{\mathcal{M}} = -\frac{\partial \Phi}{\partial \boldsymbol{\Omega}_0}.
\end{equation}
The above relations ensure the validity of the Gibbs-Duhem relation,
\begin{equation}
 d\Phi = -\mathcal{S} dT_0 - \mathcal{Q} d\mu_0 - P dV - \boldsymbol{\mathcal{M}} \cdot d\boldsymbol{\Omega}_0.
\end{equation}

To satisfy the above equations, we can compute $\Phi$ by integrating the total energy at constant $\beta_0 \mu_0$ and $\beta_0 \Omega_0$ \cite{Patuleanu:2025zbn}:
\begin{equation}
 \Phi = \frac{1}{\beta_0} \int d\beta_0\, (\mathcal{E})_{\beta_0\mu_0, \beta_0 \Omega_0}.
 \label{eq:GCE_Phi}
\end{equation}
The best candidate for the total energy is just the volume integral of the energy-momentum tensor, $\mathcal{E} = \int d^3x \, \Theta^{tt}$,\footnote{Note that the diagonal components of the energy-momentum tensor in the canonical and Belinfante [see Eq.~\eqref{eq:Belinfante}] pseudogauges are identical.} such that the grand canonical potential density takes the familiar form \cite{Jiang:2016wvv,Wang:2018sur}
\begin{equation}
 \phi(x) = -\frac{1}{8\pi^2 \beta_0} \sum_j F_j \left(J_j^+ + \frac{2\lambda_j k_j}{p_j} J_j^-\right),
 \label{eq:GCE_phi}
\end{equation}
where $F_j = \ln(1 + e^{-\beta \widetilde{\mathcal{E}}_j})$.
Writing $\int dE\, E \int dk \to \int dq\, q \int dk$ and integrating by parts under the  $k$ integral shows that 
\begin{equation}
 \phi(x) = -\Theta^{zz}.
\end{equation}
It is easy to check that 
\begin{equation}
 -\frac{\partial \phi}{\partial \mu_0} = J^t_V, \quad 
 -\frac{\partial \phi}{\partial \Omega_0} = \rho^2 \Theta^{t\varphi} + \frac{1}{2} J^z_A = M_C^{t,xy},
\end{equation}
with $M_C^{\mu,\alpha\beta}$ being the canonical angular momentum density,
\begin{equation}
 M^{\mu,\alpha\beta}_C = x^\alpha \Theta^{\mu\beta} - x^\beta \Theta^{\mu\alpha} + S_C^{\mu,\alpha\beta},
 \label{eq:ang}
\end{equation}
while $S_C^{\mu,\alpha\beta} = \frac{i}{8} \bar{\psi} \{\gamma^\mu, [\gamma^\alpha,\gamma^\beta]\} \psi = -\frac{1}{2}\varepsilon^{\mu\alpha\beta\nu} J_{A;\nu}$ corresponds to the canonical spin angular momentum density \cite{Itzykson:1980rh,Peskin:1995ev}, where we take the convention $\varepsilon^{txyz} = 1$ for the Levi-Civitta tensor. It can be seen that the total charge and angular momentum are given by $\mathcal{Q} = \int d^3x\, J^t_V$ and $\mathcal{M} = \int d^3x\, M^{t,xy}_C$. We remark that the emergence of the total angular momentum $M^{t,xy}_C$ in the canonical pseudogauge, and not in other pseudogauges, is not by explicit choice, but rather follows as a consequence of the definition of the thermodynamic potential in Eq.~\eqref{eq:GCE_phi}.

The entropy density, defined by $\mathcal{S} = \int_V d^3x S$ , is given by
\begin{equation}
 S = \frac{1}{T_0} (\Theta^{tt} - \phi - \mu_0 J^t_V - \rho^2 \Omega_0 \Theta^{t\varphi} - \tfrac{1}{2} \Omega_0 J^z_A).
 \label{eq:GCE_S_aux}
\end{equation}
Taking into account that the local four-velocity of a rigidly rotating gas is 
\begin{equation}
 u^\mu \partial_\mu = \Gamma (\partial_t + \Omega_0 \partial_\varphi), \quad \Gamma =(1-\rho^2\Omega_0^2)^{-1/2},
 \label{eq:umu}
\end{equation}
it is easy to recognize that $\Theta^{tt} - \rho^2 \Omega_0 \Theta^{t\varphi} = \Gamma^{-1} \Theta^{t\mu} u_\mu$. By virtue of the Tolman-Ehrenfest law, the local temperature and chemical potential are given by
\begin{equation}
 T = \Gamma T_0, \quad \mu = \Gamma \mu_0.
 \label{eq:Tmu_local}
\end{equation}
We can thus rewrite Eq.~\eqref{eq:GCE_S_aux} in the following form:
\begin{equation}
 S = \frac{1}{T} \left(\Theta^{t\mu} u_\mu - \Gamma \phi - \mu J^t_V - \frac{1}{2} S^{t,\alpha\beta}_C \omega_{\alpha\beta}\right).
 \label{eq:GCE_S}
\end{equation}
The vorticity tensor $\omega_{\alpha\beta}$ appearing above can be written with respect to the local acceleration and vorticity four-vectors, $a^\mu = u^\alpha\partial_\alpha u^\mu$ and $\omega^\mu = \frac{1}{2} \varepsilon^{\mu\nu\alpha\beta} u_\nu \partial_\alpha u_\beta$, as follows \cite{Becattini:2015nva}:
\begin{gather}
 \omega_{\alpha\beta} = a_\alpha u_\beta - a_\beta u_\alpha - \varepsilon_{\alpha\beta\mu\nu} u^\mu \omega^\nu, \nonumber\\
 a^\mu = \omega^{\mu\nu} u_\nu, \quad 
 \omega^\mu = -\frac{1}{2} \varepsilon^{\mu\nu\alpha\beta} u_\nu \omega_{\alpha\beta}.
 \label{eq:vorticity_tensor}
\end{gather}
For the velocity profile in Eq.~\eqref{eq:umu}, we have 
\begin{equation}
 a^\mu \partial_\mu = -\rho \Omega_0^2 \Gamma^2 \partial_\rho, \quad 
 \omega^\mu \partial_\mu = \Omega_0 \Gamma^2 \partial_z,
 \label{eq:aomega}
\end{equation}
while $\omega_{\alpha\beta} = \Omega_0 \Gamma (g_{\alpha x} g_{\beta y} - g_{\alpha y} g_{\beta x})$, such that $\frac{1}{2} S^{t,\alpha\beta}_C \omega_{\alpha\beta} = \Gamma \Omega_0 S^{t,xy}_C$, with $S^{t,xy}_C = \frac{1}{2} J^z_A$.

\subsection{Thermodynamic potential current}\label{sec:GCE:curr}

It is tempting to write Eq.~\eqref{eq:GCE_S} in covariant form by interpreting $s^t = S$ and $\tilde{\phi}^t = \Gamma \phi$ as the time components of the entropy and thermodynamic potential four-vectors. We will exploit this generalization in this section. First, we construct the quantity $\phi^\mu$ in analogy to Eq.~\eqref{eq:GCE_Phi}, i.e.
\begin{equation}
 \phi^\mu = \frac{1}{\beta_0} \int d\beta_0 (\Theta^{\mu t})_{\beta_0 \mu_0, \beta_0\Omega_0},
 \label{eq:GCE_phimu}
\end{equation}
with $\phi^t \equiv \phi$ given in Eq.~\eqref{eq:GCE_phi}, $\phi^\rho = 0$ and
\begin{align}
 \phi^\varphi &= -\frac{T_0}{8\pi^2 \rho} \sum_j F_j \frac{q_j J_j^\times}{E_j}, \nonumber\\
 \phi^z &= -\frac{T_0}{8\pi^2} \sum_j F_j \left(\frac{k_j}{E_j} J^+_j + \frac{2\lambda_j p_j}{E_j} J_j^-\right).
 \label{eq:GCE_phimu_phiz}
\end{align}
It is easy to check that 
\begin{equation}
 J^\mu_V = -\frac{\partial \phi^\mu}{\partial \mu_0}, \quad 
 M_C^{\mu,xy} = -\frac{\partial \phi^\mu}{\partial \Omega_0},
 \label{eq:GCE_phimu_JM}
\end{equation}
while the entropy current $s^\mu = -\partial \phi^\mu / \partial T_0$ reads 
\begin{align}
 s^\mu &= \frac{1}{T_0} (\Theta^{\mu t} - \phi^\mu - \mu_0 J^\mu_V - \rho^2 \Omega_0 \Theta^{\mu\varphi} + \tfrac{1}{2} \Omega_0 \varepsilon^{\mu xy \nu} J_{A;\nu}) \nonumber\\
 &= \frac{1}{T} \left(\Theta^{\mu\nu} u_\nu - \tilde{\phi}^\mu - \mu J^\mu_V - \frac{1}{2} S^{\mu,\alpha\beta}_C \omega_{\alpha\beta}\right),
 \label{eq:GCE_smu}
\end{align}
where we identified the local thermodynamic current,
\begin{equation}
 \tilde{\phi}^\mu = \Gamma \phi^\mu.
 \label{eq:phit}
\end{equation}
Contracting Eq.~\eqref{eq:GCE_smu} with the fluid four-velocity $u_\mu$ reveals
\begin{equation}
 s = \frac{1}{T} \left(\epsilon + P - \mu Q_{V} - \frac{1}{2} S^{\alpha\beta}_C \omega_{\alpha\beta}\right),
 \label{eq:GCE_sdens}
\end{equation}
where $\epsilon = u_\mu \Theta^{\mu\nu}u_\nu$ is the energy density, $Q_V = u_\mu J^\mu_V$ and $S^{\alpha\beta}_C = u_\mu S^{\mu,\alpha\beta}_C$ are the charge and spin densities, while $P = -\tilde{\phi}^\mu u_\mu$ represents the thermodynamic pressure. 

We stress that all relations derived in this section are exact and thermodynamically consistent. However, before interpreting Eq.~\eqref{eq:GCE_sdens} as a local thermodynamic relation, we must shift the paradigm from the global canonical ensemble to the local thermodynamic state of the system, as discussed in the following section. Our approach differs from that of Refs.~\cite{Becattini:2025oyi,Becattini:2019poj,Rindori:2021quq}, where the thermodynamic potential current is constructed by subtracting the vanishing temperature limit to achieve thermodynamic consistency. In our approach, such a subtraction is not necessary.

\section{Local state thermodynamics}\label{sec:local}

The local state of the rotating fluid is characterized by the local temperature and chemical potential given in Eq.~\eqref{eq:Tmu_local}. Furthermore, the non-inertial motion of the fluid gives rise to a kinematic tetrad comprised of the fluid velocity $u^\mu$, the acceleration $a^\mu$ and vorticity $\omega^\mu$ vectors given in Eq.~\eqref{eq:aomega}, and fourth vector $\tau^\mu = \varepsilon^{\mu\nu\alpha\beta} u_\nu a_\alpha \omega_\beta$, explicitly given by
\begin{gather}
 \tau^\mu \partial_\mu = -\Omega_0^3 \Gamma^5 (\rho^2 \Omega_0 \partial_t + \partial_\varphi).
 \label{eq:tau}
\end{gather}

To account for the vortical effects locally, the canonical set $\{T, \mu\}$ of thermodynamic parameters must be extended. In this paper, we consider an extension by the spin potential, $\Omega^{\mu\nu}$, which is known to relax to the  vorticity $\omega^{\mu\nu}$ in thermal equilibrium \cite{Wagner:2024rbt,Wagner:2024fry}. Decomposing the spin tensor by analogy to Eq.~\eqref{eq:vorticity_tensor}, \cite{Becattini:2025oyi,Florkowski:2017ruc,Ambrus:2022yzz,Sapna:2025yss}
\begin{gather}
 \Omega_{\alpha\beta} = \kappa_\alpha u_\beta - \kappa_\beta u_\alpha - \varepsilon_{\alpha\beta\mu\nu} u^\mu \Omega^\nu, \nonumber\\
 \kappa^\mu = \Omega^{\mu\nu} u_\nu, \quad 
 \Omega^\mu = -\frac{1}{2} \varepsilon^{\mu\nu\alpha\beta} u_\nu \Omega_{\alpha\beta},
\end{gather}
we identify its electric (acceleration) and magnetic (vortical) components, $\kappa^\mu$ and $\Omega^\mu$. The goal of this section is to provide a formulation of the vortical effects, in which certain instances of the vorticity $\omega^\mu$ and acceleration $a^\mu$ are replaced by the equivalent spin-potential quantities, $\kappa^\mu$ and $\Omega^\mu$, such that the system exhibits local thermodynamic consistency.

\subsection{From global ensemble to local state}\label{sec:local:local}

The transition $(T_0, \mu_0, \Omega_0) \to (T,\mu,\omega_{xy}) = (\Gamma T_0, \Gamma \mu_0, \Gamma \Omega_0)$ from global (GCE) to local parameters arises naturally, since the distribution function $f_j$ \eqref{eq:fj} can be written as
\begin{equation}
 f_j = \left[e^{\beta (\Gamma E_j - \sigma_j \mu - \omega_{xy} m_j) } + 1\right]^{-1}.
 \label{eq:fj_loc}
\end{equation}
The first step towards the local formulation is to treat $\tilde{\phi}^\mu = \Gamma \phi^\mu \equiv \tilde{\phi}^\mu(T, \mu, \omega_{xy})$ as a function just of local quantities. Employing the relations in Eq.~\eqref{eq:GCE_phimu_JM}--\eqref{eq:GCE_smu}, it can be seen that 
\begin{equation}
 \frac{\partial \tilde{\phi}^\alpha}{\partial T} = \frac{\partial \phi^\alpha}{\partial T_0} = -s^\alpha, \quad 
 \frac{\partial \tilde{\phi}^\alpha}{\partial \mu} = \frac{\partial \phi^\alpha}{\partial \mu_0} = -J^\alpha_V.
 \label{eq:local_dphi}
\end{equation}
Taking now the derivative $\partial \phi^\mu / \partial \Omega_0$, we arrive at
\begin{align}
 \frac{\partial \phi^\mu}{\partial \Omega_0} 
 &= \Gamma^2 \frac{\partial \tilde{\phi}^\mu}{\partial \omega_{xy}} - \rho^2 \Omega_0 \Gamma (\tilde{\phi}^\mu - T s^\mu - \mu J^\mu_V).
\end{align}
Rearranging the above terms leads to
\begin{equation}
 -\frac{\partial \tilde{\phi}^\mu}{\partial \omega_{xy}} = S_C^{\mu,xy} + \Theta^{\mu\nu} \tilde{\tau}_\nu,
 \label{eq:dphit_domega}
\end{equation}
where $\tilde{\tau}^\mu \partial_\mu = -\rho^2 \Omega_0 \partial_t - \partial_\varphi$ is a four-vector orthogonal to the four-velocity $u^\mu$. The term $\Theta^{\mu\nu} \tilde{\tau}_\nu$ is not completely unexpected: it appears because in our formalism, the parameter $\Omega_0$ is responsible both for orbital effects, which we associate with the $\Theta^{\mu\nu} \tilde{\tau_\nu}$ term, and for genuine quantum vortical effects, which we wish to associate with the spin tensor contribution. 

To disentangle kinematic, orbital contributions from the local, spin contributions, we have to consider that $\tilde{\phi}^\mu$ is a function of the spin potential $\Omega_{\alpha\beta}$, instead of the vorticity tensor $\omega_{\alpha\beta}$. In thermal equilibrium, these two quantities are identical and therefore, the derivative in Eq.~\eqref{eq:dphit_domega} is comprised of both the kinematic and the spin parts, which satisfy individually
\begin{equation}
 -\frac{\partial \tilde{\phi}^\mu}{\partial \omega_{xy}} = \Theta^{\mu\nu} \tilde{\tau}_\nu, \quad 
 -\frac{\partial \tilde{\phi}^\mu}{\partial \Omega_{xy}} = S^{\mu,xy}_C.
\end{equation}
At the level of the general formulas considered so far, the split illustrated above seems untractable. For this reason, we will consider in the following subsection the special case of massless fermions, where analytic expressions are available.

\subsection{Massless fermions: local thermodynamic potential current}\label{sec:local:m0}

In the case of massless fermions, the components of the energy-momentum tensor and charge currents are known analytically \cite{Ambrus:2019khr}. With respect to the so-called beta frame velocity \cite{Van:2013gna,Becattini:2014yxa}, considered in Eq.~\eqref{eq:umu}, the energy-momentum tensor and the vector and axial currents can be decomposed as
\begin{gather}
 T^{\mu\nu} = \epsilon u^\mu u^\nu - P_{\rm eff} \Delta^{\mu\nu} + \pi^{\mu\nu} + \sigma^\tau_\varepsilon(\tau^\mu u^\nu + \tau^\nu u^\mu), \nonumber\\
 J^\mu_V = Q_V u^\mu + \sigma^\tau_V \tau^\mu, \quad 
 J^\mu_A = \sigma^\omega_A \omega^\mu,
 \label{eq:m0_Tmunu}
\end{gather}
where $T^{\mu\nu}$ is the energy-momentum tensor in the Belinfante pseudogauge. Full derivation details for the quantities appearing above are available in Ref.~\cite{Ambrus:2019khr}, for the case of massless fermions in the presence of the vector $\mu_V$, axial $\mu_A$ and helical $\mu_H$ chemical potentials. Here, we considered the case $\mu_A = \mu_H = 0$, when the shear stress tensor reads
\begin{equation}
 \pi^{\mu\nu} = -\frac{2}{27\pi^2} \left(\tau^\mu \tau^\nu + \frac{\omega^2}{2} a^\mu a^\nu + \frac{a^2}{2} \omega^\mu \omega^\nu \right),
\end{equation}
with $\omega^2 = \omega_\mu \omega^\mu = -\Omega_0^2 \Gamma^4$ and $a^2 = a_\mu a^\mu = -\Omega_0^2 \Gamma^2(\Gamma^2 - 1)$, while the scalar quantities in Eq.~\eqref{eq:m0_Tmunu} read
\begin{gather}
 P_{\rm eff} = P_{\rm cl} - \frac{3\omega^2 + a^2}{12} \sigma^\omega_{A;{\rm cl}} + \frac{\omega^4 + \tfrac{46}{45} \omega^2 a^2 - \tfrac{17}{15} a^4}{192\pi^2}, \nonumber\\
 \sigma^\tau_\varepsilon = -\frac{1}{3} \sigma^\omega_{A;{\rm cl}} + \frac{39 \omega^2 + 31 a^2}{360 \pi^2},\quad 
 \sigma^\omega_A = \sigma^\omega_{A;{\rm cl}} - \frac{\omega^2 + 3 a^2}{24\pi^2}, \nonumber\\
 Q_V = Q_{V;{\rm cl}} - \frac{\mu (\omega^2 + a^2)}{4\pi^2}, \quad 
 \sigma^\tau_V = \frac{\mu}{6\pi^2}. 
 \label{eq:m0_QFT_results}
\end{gather}
The energy density satisfies $\epsilon = 3P_{\rm eff}$ and $P_{\rm eff} = P + \Pi$ is the effective pressure, comprised of the thermodynamic and dynamic pressures, $P$ and $\Pi$. The classical contributions given above read
\begin{equation}
 P_{\rm cl} = \frac{7\pi^2 T^4}{180} + \frac{\mu^2 T^2}{6} + \frac{\mu^4}{12\pi^2}, \quad 
 \sigma^\omega_{A;{\rm cl}} = \frac{T^2}{6} + \frac{\mu^2}{2\pi^2}, 
 \label{eq:Pcl}
\end{equation}
with $Q_{V;{\rm cl}} = \partial P_{\rm cl} / \partial \mu = \mu T^2 / 3 + \mu^3 / 3\pi^2$. The results in Eqs.~\eqref{eq:m0_Tmunu}--\eqref{eq:Pcl} are consistent with those derived also in Refs.~\cite{Buzzegoli:2017cqy,Buzzegoli:2018wpy,Prokhorov:2018bql,Prokhorov:2019yft,Palermo:2021hlf}.

We now use the above expressions to find the thermodynamic potential current, $\tilde{\phi}^\mu = \Gamma \phi^\mu$, by employing Eq.~\eqref{eq:GCE_phimu}. To employ this equation, we must find the components of the canonical energy-momentum tensor, $\Theta^{\mu\nu}$, which is related to the Belinfante energy-momentum tensor $T^{\mu\nu}$ via a pseudogauge transformation employing the superpotential equal to the canonical spin tensor \cite{Florkowski:2018fap},
\begin{equation}
 T^{\mu\nu} = \Theta^{\mu\nu} +\frac{1}{2} \partial_\lambda(S_C^{\lambda,\mu\nu} + S_C^{\mu,\nu\lambda} - S_C^{\nu,\lambda\mu}),
 \label{eq:Belinfante}
\end{equation}
where $\nabla_\lambda$ represents the covariant derivative,
which leads to $\Theta^{t\varphi} = T^{t\varphi} + \frac{1}{4\rho} \partial_\rho J^z_A$ and $\Theta^{\varphi t} = T^{\varphi t} - \frac{1}{4\rho} \partial_\rho J^z_A$. Using $\tilde{\phi}^\mu = \frac{\Gamma}{\beta_0} \int d\beta_0 \, \Theta^{\mu t}\rvert_{\beta_0 \mu_0, \beta_0 \Omega_0}$ leads to:
\begin{align}
 \tilde{\phi}^t &= -\Gamma \left[P_{\rm cl} - \frac{3\omega^2 + a^2}{12}  \sigma^\omega_{A;{\rm cl}} + \frac{\omega^4 + \tfrac{122}{15} \omega^2 a^2 - \tfrac{17}{15} a^4}{192 \pi^2} \right], \nonumber\\
 \tilde{\phi}^\varphi &= -\Omega_0 \Gamma \left[P_{\rm cl} - \frac{\omega^2 + 3 a^2}{12} \sigma^\omega_{A;{\rm cl}} + \frac{\omega^4 + 22 \omega^2 a^2 + 17 a^4}{960\pi^2}\right],
\end{align}
while $\tilde{\phi}^\rho = \tilde{\phi}^z = 0$. 

With respect to the kinematic tetrad in Eqs.~\eqref{eq:aomega} and \eqref{eq:tau}, the thermodynamic potential vector can be written as 
\begin{equation}
 \tilde{\phi}^\mu = -P u^\mu - \sigma_\phi^\tau \tau^\mu, 
\end{equation}
where the thermodynamic pressure $P = -u_\mu \tilde{\phi}^\mu$ and circular conductivity $\sigma_\phi^\tau = -\tau_\mu \tilde{\phi}^\mu / \tau^2$ read
\begin{align}
 P &= P_{\rm cl} - \frac{\omega^2 + a^2}{4} \sigma^\omega_{A;{\rm cl}} 
 + \frac{\omega^4 + \tfrac{134}{15} \omega^2 a^2 + \tfrac{17}{5} a^4}{192\pi^2}, \nonumber\\
 \sigma^\tau_\phi &= \frac{1}{6} \sigma^\omega_{A;{\rm cl}} - \frac{3 \omega^2 + 17 a^2}{720\pi^2}.
 \label{eq:Pth}
\end{align}
The thermodynamic pressure above satisfies the Euler relation in Eq.~\eqref{eq:GCE_sdens}, where $\epsilon = 3 P_{\rm eff}$ is obtained from Eq.~\eqref{eq:m0_QFT_results}, while $s = \partial P / \partial T$ and $Q_V = \partial P / \partial \mu$ read
\begin{equation}
 s = s_{\rm cl} - \frac{T(\omega^2 + a^2)}{12}, \quad 
 Q_V = Q_{V;{\rm cl}} - \frac{\mu(\omega^2 + a^2)}{4\pi^2},
\end{equation}
with $s_{\rm cl} = \partial P_{\rm cl} /\partial T = 7\pi^2 T^3 / 45 + \mu^2 T / 3$. Moreover, the dynamic pressure can be obtained as $\Pi = P_{\rm eff} - P$,
\begin{equation}
 \Pi = \frac{a^2}{6} \sigma^\omega_{A;{\rm cl}} - \frac{a^2 (89 \omega^2 + 51 a^2)}{2160\pi^2}.
\end{equation}

\subsection{Local spin potential}\label{sec:local:spinpot}

As shown in Eq.~\eqref{eq:Pth}, the thermodynamic pressure depends on local vorticity and acceleration. We now assume that, out of equilibrium, the pressure depends partially on these kinematic quantities, and partially on the quantities related to the spin potential. The derivatives of $\Omega^2 = \Omega_\lambda \Omega^\lambda$ and $\kappa^2 = \kappa_\lambda \kappa^\lambda$ with respect to the spin potential $\Omega_{\alpha\beta}$ can be obtained as
\begin{equation}
 \frac{\partial \Omega^2}{\partial \Omega_{\alpha\beta}} = 2(u^\beta \kappa^\alpha - u^\alpha \kappa^\beta - \Omega^{\alpha\beta}), \quad 
 \frac{\partial \kappa^2}{\partial \Omega_{\alpha\beta}} = 2 (u^\beta \kappa^\alpha - u^\alpha \kappa^\beta).
\end{equation}

We now turn to the spin angular momentum density tensor $S^{\mu,\alpha\beta}_C = -\frac{1}{2} \varepsilon^{\mu\alpha\beta\nu} J_{A;\nu}$, introduced in Eq.~\eqref{eq:ang}. As shown in Eq.~\eqref{eq:m0_Tmunu}, $J^\mu_A = \sigma^\omega_A \omega^\mu$, with $\omega^\mu$ being the vorticity vector introduced in Eq.~\eqref{eq:vorticity_tensor}.
We now assume that, in the local description of the fluid, the spin density $S_C^{\alpha\beta} = u_\mu S^{\mu,\alpha\beta}_C$ remains proportional to the vorticity tensor. This is compatible with the results reported in Ref.~\cite{Singha:2025bda}, where it was shown that the expectation value of the canonical spin operator at finite spin potential vanishes for massless fermions. We thus seek to obtain:
\begin{equation}
 S_C^{\alpha\beta} = \frac{1}{2} (\omega^{\alpha\beta} + u^\alpha a^\beta - u^\beta a^\alpha) \sigma^\omega_A = \frac{\partial P}{\partial \Omega_{\alpha\beta}},
 \label{eq:Sab_from_P}
\end{equation}
where we keep for simplicity $\sigma_A^\omega$ independent of the spin potential $\Omega_{\alpha\beta}$, as given in Eq.~\eqref{eq:m0_QFT_results}. 

The tensor structure of $S^{\alpha\beta}_C$ is compatible with the derivative of the magnetic part $\Omega^\mu$ of the spin potential, and incompatible with that of its electric part, $\kappa^\mu$. 
Imposing that $\partial P / \partial \Omega_{\alpha\beta} = S^{\alpha\beta}_C$ implies that the thermodynamic pressure $P$ in Eq.~\eqref{eq:Pth} should be enhanced with the following term: 
\begin{multline}
 P = P_{\rm cl} - \frac{\omega^2 + a^2}{4} \sigma^\omega_{A;{\rm cl}} 
 + \frac{\omega^4 + \tfrac{134}{15} \omega^2 a^2 + \tfrac{17}{5} a^4}{192\pi^2} \\
 - \frac{1}{2} (\omega \cdot \delta \Omega) \left(\sigma^\omega_{A;{\rm cl}} - \frac{\omega^2 + 3a^2}{24\pi^2}\right),
 \label{eq:final_P}
\end{multline}
where $\delta \Omega^\mu= \Omega^\mu - \omega^\mu$ represents the deviation of the spin potential vector from the vorticity vector.
The above form of the pressure is inspired by the structure of $S^{\alpha\beta}_C$, which motivates the structure of the terms proportional to $\omega \cdot \delta \Omega$.
Taking the derivatives of $P$ with respect to $T$, $\mu$ and $\Omega_{\alpha\beta}$ reveals the thermodynamic relations 
\begin{gather}
 s = s_{\rm cl} -\frac{T(\omega^2 + a^2)}{12} - \frac{T}{6} \omega \cdot \delta \Omega, \quad \nonumber \\
 Q_V = Q_{V;{\rm cl}} - \frac{\mu(\omega^2 + a^2)}{4\pi^2} - \frac{\mu}{2\pi^2} \omega \cdot \delta \Omega,
 \nonumber\\
 \frac{\partial P}{\partial \Omega_{\alpha\beta}} = S^{\alpha\beta}_C = \frac{1}{2}(\omega^{\alpha\beta} + u^\alpha a^\beta - u^\beta a^\alpha) \sigma_A^{\omega}.
 \label{eq:final_sQsigma}
\end{gather}
To find the energy density, the Euler relation \eqref{eq:GCE_sdens} is modified to 
$\epsilon = Ts -P + \mu Q_V + \frac{1}{2} \Omega_{\alpha\beta} S^{\alpha\beta}_C$, leading to
\begin{equation}
 \epsilon = \epsilon_{\rm cl}-\left(\frac{3 \omega^2+a^2}{4} + \omega \cdot \delta \Omega\right) \sigma_{A;cl}^{\omega}+\frac{\omega^4 + \frac{46}{45} \omega^2 a^2 -\frac{17}{15} a^4}{64\pi^2}.
 \label{eq:final_eps}
\end{equation}
Knowing that $P_{\rm eff} = \epsilon / 3 = P + \Pi$, we can identify the dynamic pressure as
\begin{equation}
 \Pi = \frac{a^2 + \omega \cdot \delta \Omega}{6} \sigma^\omega_{A;{\rm cl}} - \frac{a^2 (89 \omega^2 + 51 a^2)}{2160\pi^2} 
 - \frac{3 a^2 + \omega^2}{48\pi^2} \omega \cdot \delta \Omega.
 \label{eq:final_Pi}
\end{equation}
We are now in a position to write the thermodynamic potential current $\tilde{\phi}^\mu$, by solving 
\begin{equation}
 S^{\mu,\alpha\beta}_C = -\frac{1}{2} \sigma^\omega_A \varepsilon^{\mu\alpha\beta\lambda} \omega_\lambda = -\frac{\partial \tilde{\phi}^\mu}{\partial\Omega_{\alpha\beta}}.
\end{equation}
The result can be expressed as
\begin{equation}
 \tilde{\phi}^\mu = -P u^\mu - \sigma^\tau_\phi \tau^\mu - \delta \sigma^\tau_\phi \delta \tau^\mu,
 \label{eq:final_phimu}
\end{equation}
where $P$ is the modified pressure in Eq.~\eqref{eq:final_P}, the circular conductivity $\sigma^\tau_\phi$ is given in Eq.~\eqref{eq:Pth}, $\delta \sigma^\tau_\phi = \frac{1}{2} \sigma^\omega_A$, while the four-vectors $\tau^\mu$, defined before Eq.~\eqref{eq:tau}, and $\delta \tau^\mu$ are given as
\begin{equation}
 \tau^\mu = \varepsilon^{\mu\nu\alpha\beta} u_\nu a_\alpha \omega_\beta, \quad 
 \delta \tau^\mu = \varepsilon^{\mu\nu\alpha\beta} u_\nu (\kappa_\alpha - a_\alpha) \omega_\beta.
\end{equation}

Eqs.~\eqref{eq:final_P}, \eqref{eq:final_sQsigma}, \eqref{eq:final_eps}, \eqref{eq:final_Pi} and \eqref{eq:final_phimu} represent the main result of this paper. 
In our proposed framework, the energy density $\epsilon = u_\mu \Theta^{\mu\nu} u_\nu$, spin density $S^{\alpha\beta}_C = u_\mu S_C^{\mu,\alpha\beta}$ and charge density $Q_V = u_\mu J^\mu_V$ are computed from the local energy-momentum tensor, spin tensor and charge current, while the entropy density $s$ is given by the Euler relation \eqref{eq:final_eps}. Furthermore, $s$, $Q_V$ and $S^{\alpha\beta}_C$ are related to the usual derivatives of the thermodynamic pressure $P$ via Eqs.~\eqref{eq:final_sQsigma}.

We end this section by remarking that our results are seemingly in contradiction with those in Refs.~\cite{Becattini:2023ouz,Becattini:2025oyi}, where it is claimed that the Euler relation \eqref{eq:final_eps} and the differential relations \eqref{eq:final_sQsigma} cannot be satisfied simultaneously. The two main differences between our approach and that in Refs.~\cite{Becattini:2023ouz,Becattini:2025oyi} are that: 1) we allow the pressure to be a function of both the spin potential and the vorticity tensor; and 2) we allow the system to develop a dynamic pressure, absorbing the difference between the effective pressure $P_{\rm eff} = \frac{1}{3} \epsilon$ and the thermodynamic pressure $P$.

\section{Conclusion}\label{sec:conc}

In this work, we considered the thermodynamics of a rigidly rotating gas of Dirac particles. Starting from the grand canonical ensemble (GCE), defining a system rotating with constant angular velocity $\Omega_0$, at temperature $T_0$ and chemical potential $\mu_0$, we identified the grand canonical potential \eqref{eq:GCE_phi} that allows to extract the total entropy, charge and angular momentum, contained in a fictitious cylinder of macroscopic radius $R$ (see Sec.~\ref{sec:GCE:GP}). 

Within the GCE, we constructed the thermodynamic potential current, $\phi^\mu$ \eqref{eq:GCE_phimu_phiz}, which allows the entropy and charge currents to be recovered using usual thermodynamic relations. The derivative with respect to the angular velocity $\Omega_0$ returns the total angular momentum density in the canonical pseudogauge [see Eq.~\eqref{eq:GCE_phimu_JM}].

We further discussed the transition from the parameters of the GCE to the parameters defining the local state of the fluid: the local temperature, chemical potential, vorticity tensor and spin potential. We showed in Eq.~\eqref{eq:final_P} that, away from equilibrium, the thermodynamic pressure receives corrections due to acceleration, vorticity and spin potential. The resulting thermodynamic pressure satisfies both the Euler relation and the thermodynamic relations with respect to the entropy density, charge density and spin density [see Eq.~\eqref{eq:final_sQsigma}]. We also obtained the covariant thermodynamic potential current $\tilde{\phi^{\mu}}$~\eqref{eq:final_phimu}.

Our findings provide a firm, quantum-field-theoretical grounding of spin hydrodynamics, providing a recipe for formulating the dynamics of fluids with spin in a thermodynamically consistent way, which is compatible with the known vortical effects derived in quantum field theory. 
Furthermore, our results may be applicable to the wide range of studies involving rigid rotation, such as phase diagram of strongly interacting matter under rotation \cite{Chernodub:2016kxh,Zhu:2025pxh}, dilepton production in rotating quark-gluon plasma \cite{Castano-Yepes:2025zae} or spin polarization in heavy-ion collisions \cite{Palermo:2023cup}.\vspace{-10pt}

\section*{Acknowledgements} \vspace{-5pt}
The authors are grateful to Dr. David Wagner, Dr. Nyx Shiva, Dr. Matteo Buzzegoli, Dr. Andrea Palermo and Dr. Maxim Chernodub for useful comments on the manuscript. This work was funded by the EU’s NextGenerationEU instrument through the National Recovery and Resilience Plan of Romania - Pillar III-C9-I8, managed by the Ministry of Research, Innovation and Digitization, within the project entitled ``Facets of Rotating Quark-Gluon Plasma'' (FORQ), contract no.~760079/23.05.2023 code CF 103/15.11.2022.

\bibliographystyle{elsarticle-num}
\bibliography{plasma}

\end{document}